\documentstyle[aps,prbbib,epsf,times]{revtex}

\begin{document}
\draft
\tighten

\wideabs{ 
\title{Superconducting transitions from the pseudogap
state: \textit{d}-wave symmetry, lattice,\\ and low-dimensional effects }

\author{Qijin Chen, Ioan Kosztin, Boldizs\'ar Jank\'o, and
K. Levin} 

\address{The James Franck Institute, University of
Chicago, 5640 South Ellis Avenue, Chicago, Illinois 60637}

\date{May 4, 1998}

\maketitle

\begin{abstract}
  We investigate the behavior of the superconducting transition
temperature within a previously developed BCS-Bose Einstein crossover
picture.  This picture, based on a decoupling scheme of
Kadanoff and Martin, further extended by Patton, can be used to derive
a simple form for the superconducting transition temperature in the
presence of a pseudogap.  We extend previous work which addressed the
case of $s$-wave pairing in jellium, to explore the solutions for
$T_c$ as a function of variable coupling in more physically relevant
situations. We thereby ascertain the effects of reduced
dimensionality, periodic lattices and a $d$-wave pairing interaction.
Implications for the cuprate superconductors are discussed.
\end{abstract}

\pacs{PACS numbers: 
74.20.-z, 
%
%
74.20.Fg, 
%
74.25.Dw, 
%
74.25.Jb
%
%
\hfill PRB \textbf{59}, 7083 (1999).\qquad  \textsf{\textbf{cond-mat/9805032}}
}
}

\section{Introduction}
\label{Section_1}

The concept of a smooth evolution from a BCS description of
superconductivity to that of Bose Einstein condensation (BEC) dates back to
Eagles\cite{Eagles} and to Leggett.\cite{Leggett} The latter addressed this
problem at zero temperature in the context of $p$-wave pairing in He$^3$.
Nozieres and Schmitt-Rink\cite{NSR} (NSR) extended Leggett's formalism to
calculations of $T_c$ and, for the case of a jellium gas, found a
continuous variation from the BCS exponential dependence (on coupling
 constant $g$) to the Bose-Einstein asymptote, at large $g$.
Uemura\cite{Uemura} and, independently, Randeria\cite{Randeria} and
Micnas\cite{Micnas} and their respective co-workers applied this BCS
Bose-Einstein crossover picture to the high-temperature superconductors,
which, because of their short coherence length, were claimed to correspond
to intermediate values of the coupling.  It was, subsequently, argued by
these and other groups that the pseudogap (normal) state of the cuprates was
naturally associated with this intermediate coupling regime. Since then, a
large number of papers\cite{Haussmann,Ranninger,crossoverothers} have been
written on the crossover problem and the related pseudogap state.

The application of these crossover theories to the cuprates is made
complex by a number of important factors which involve quasi-two
dimensionality, lattice periodicity, and, finally, the introduction of
$d$-wave symmetry in the pairing interaction. It is the goal of the
present paper to discuss these three effects in the context of a
many-body-theoretic approach to the crossover problem, based on earlier
work by Kadanoff and Martin,\cite{Kadanoff} and extended by
Patton.\cite{Patton} A major advantage of this scheme is that it reduces
to BCS theory in the limit of weak coupling.  Our principal
contributions\cite{Janko,Maly2} to the larger body of work on the
BCS-Bose Einstein crossover have been based on $s$-wave pairing in
three-dimensional (3d) jellium. In the context of this diagrammatic approach
we have established that (i) there is a breakdown (for $T<T^*$) of the
Fermi liquid at intermediate $g$, which roughly coincides with the onset
of long lived (i.e., $\it{resonant}$) pairs and (ii) that this breakdown
has characteristic pseudogap features, such as a depressed density of
states at Fermi energy $E_F$, as well as a two-peaked spectral function.
The pseudogap amplitude $\Delta_{\text{pg}}$ can, moreover, be
quantified. (iii) We have examined the superconducting instability
associated with this pseudogap state and determined the three
self-consistent, coupled equations which must be satisfied for the
inter-related quantities $T_c$, $\mu$, and $\Delta_{\text{pg}}(T_c)$.
(iv) In the process we have presented a quantitative phase diagram for
$T_c$ and $T^*$ as a function of the coupling $g$.

Various aspects of low dimensionality, lattice effects, and non-$s$-wave
pairing interactions have been addressed in the literature within a
crossover scenario.  Schmitt-Rink, Varma, and Ruckenstein\cite{RVS}
applied the NSR approach to two-dimensional (2d) systems and found a
breakdown of the Fermi liquid even for arbitrarily weak coupling $g$.
This was manifested as a negative chemical potential $\mu$, which
occurred in conjunction with $T_c=0$. Serene\cite{SereneNSR} suggested
that this breakdown was an artifact of the NSR scheme, which is not
conserving.\cite{Balseirotried_to_correctin2d} Yamada and
co-workers\cite{Yamada} introduced a diagrammatic ``mode-mode coupling"
scheme, following similar work on related magnetic problems, and they
found that $\mu$ was properly positive at weak coupling, while $T_c$
remained zero, as expected. However, they were unable to find a
continuous crossover between the limits of very strong and weak
interactions.

Lattice effects were discussed in the original Nozieres and Schmitt-Rink
paper (and later by Belkhir and Randeria\cite{Randeria92}). These
authors noted that the $T_c$ calculations of the jellium case, could not
be readily extended to the lattice, at least in the strong coupling
regime.  This difficulty came from a variety of issues, among which, it
was claimed, was the neglect of interactions between pairs of composite
fermions.  In addition there is a reduction in the effective pair
hopping matrix element or kinetic energy. These combined effects
conspire so that $T_c$ is expected to vanish at arbitrarily large $g$,
in contrast to the transition in the Bose-Einstein ideal gas.
Demonstrations of this effect, in the negative $U$ Hubbard model, came
later, through Monte Carlo studies,\cite{numerics} as well as from
additional analytic work based on the coherent potential
approximation.\cite{Gyorffy}

The role of non-$s$-wave pairing symmetry was initially
investigated by Leggett and, subsequently, by other groups\cite{Duan}
which addressed jellium models. It was noted\cite{Leggett} that in the
strong coupling limit  the effects of an anisotropic order parameter
are absent in the excitation spectrum, so that there can be no gap
nodes in the bosonic regime. \textit{d}-wave pairing on a lattice is not yet
amenable to Monte Carlo approaches because of the fermion sign
problem. There have been, nevertheless, some numerical solutions which
address electronic spectral functions, based on the ``fluctuation
exchange approximation" (FLEX) for the lattice $d$-wave
case.\cite{Engelbrecht} Because these were applied to a strictly two
dimensional model there was no discussion of the behavior of $T_c$ in
the more general context of the crossover problem.

The results of the present paper, in which $T_c$ is computed as a
function of $g$, can be put into the context of this background
literature.  Here we arrive at a crossover scheme which in the strict 2d
limit yields $T_c = 0 $ and in which $\mu$ smoothly interpolates from
$E_F$ to large negative values in the jellium case, as the coupling
varies from weak to strong.  Lattice effects yield a vanishing strong
coupling limit for $T_c$, associated with a reduction in the effective
kinetic energy of the bosons or pairs. The $d$-wave case on a lattice is
found to be different from the $s$-wave lattice case in one significant
respect: we find that superconductivity disappears at relatively smaller
values of $g$, as a result of $d$-wave symmetry; the pair size cannot be
less than a lattice spacing so that the pairs interact more
strongly. As a consequence their mobility is suppressed.  Consequently,
the bosonic regime is essentially never reached for the
$d$-wave case.\cite{low-n} The implications of this observation and other
aspects of these calculations for the underdoped cuprates are briefly
addressed at the end of this paper.

\section{Theoretical Framework}
\label{sec:Theory}

We consider a generic system of fermions characterized by an effective,
short range pairing interaction with Hamiltonian

\begin{eqnarray}
{\cal H} & = & \sum_{{\bf k}\sigma} \epsilon_{\bf k}
c^{\dag}_{{\bf k}\sigma} c^{\ }_{{\bf k}\sigma}
\nonumber \\
& & + \sum_{\bf k k' q} V_{\bf k, k'} 
c^{\dag}_{{\bf k}+{\bf q}/2\uparrow} 
c^{\dag}_{-{\bf k}+{\bf q}/2\downarrow} 
c^{\ }_{-{\bf k'}+{\bf q}/2\downarrow} 
c^{\ }_{{\bf k'}+{\bf q}/2\uparrow},
\label{Hamiltonian}
\end{eqnarray}
where $c^\dagger_{{\bf k}\sigma}$ creates a particle in the momentum state
${\bf k}$ with spin $\sigma $, and $\epsilon_{\mathbf{k}}$ is the energy
dispersion measured from the chemical potential $\mu$ (we take
$\hbar=k_B=1$).  For simplicity, we assume a separable pairing interaction
$V_{\bf k,k'} = g \varphi_{\bf k}\varphi_{\bf k'}$, where $g=-|g|$ is the
coupling strength; the momentum dependence of the function
$\varphi_{\mathbf{k}}$, which reflects the pairing anisotropy, will be
specified below.

In order to establish notation and to make clear our approximations, in what
follows we provide a brief description of our formalism. This approach is
based on the correlation function (or Greens function equation of motion)
formulation of the ``pairing approximation'' originally discussed by
Kadanoff and Martin\cite{Kadanoff} and later extended by
Patton.\cite{Patton}
The system can be characterized by the one- and two-particle Green's
functions which, in real space, obey the following equations:\cite{Kadanoff}

\begin{mathletters}
  \label{eq:t1}
\begin{eqnarray}
  \label{eq:t1a}
  G(1-1') &&\;=\; G_0(1-1') \\
  &&+\int\!\!d\overline{1}\,d\overline{2}\, 
  G_0(1-\overline{1})\,V(\overline{1}-\overline{2})\,
  G_2(\overline{1}\,\overline{2};1'\overline{2}^+)\;,\nonumber
\end{eqnarray}
\begin{eqnarray}
  \label{eq:t1b}
  C_2(1\,2;1'\,2') &=&
  -\int\!\!d\overline{1}\,d\overline{2}\,G(1-\overline{1})\,
  G_0(2-\overline{2})\\
  &&\times V(\overline{1}-\overline{2})\,
  G_2(\overline{1}\,\overline{2};1'\,2')\;,\nonumber
\end{eqnarray}
where the two-particle correlation function $C_2$ is given by

\begin{equation}
  \label{eq:t1c}
  C_2(1\,2;1'\,2') \;=\; G_2(1\,2;1'\,2') - 
  G(1-1')\,G_0(2-2')
\end{equation}
\end{mathletters}
and, for brevity, we have used a four vector notation $1\equiv
({\mathbf r},\tau)$, etc. While Eq.~(\ref{eq:t1a}) is exact,
Eqs.~(\ref{eq:t1b}) and (\ref{eq:t1c}) are approximate; these equations
were originally proposed by Kadanoff and Martin \cite{Kadanoff,cons_laws}
[see their Eqs.~(2.6) and, most importantly, (2.29)] as a simple decoupling
scheme for the three-particle Green's function.

This approach is primarily motivated by the observation that it yields the
results of BCS theory, in the weak coupling limit.
It is convenient to express the correlation function $C_2$ in terms of a
T-matrix (or \textit{pair propagator}) via the definition

\begin{eqnarray}
  \label{eq:t2}
  C_2(1\,2;1'\,2') &=& \int\!\!d\overline{1}\,d\overline{2}\,
  d\overline{1}'\,d\overline{2}'\, G(1-\overline{1})\,
  G_0(2-\overline{2})\\
  &&\times t(\overline{1}\,\overline{2};\overline{1}'\,\overline{2}')\,
  G(\overline{1}'-1')\,G_0(\overline{2}'-2')\;. \nonumber
\end{eqnarray}
Taking the Fourier transform of Eqs.~(\ref{eq:t1}) and (\ref{eq:t2}), and
noticing that for our separable interaction the T-matrix can be written as
$t(K,K';Q) = t(Q)\varphi_{\bf k}\varphi_{\bf k'}$, after some straightforward
algebra we obtain the following equations \cite{Patton} for the self-energy

\begin{mathletters}
  \label{eq:t3}
\begin{eqnarray}
  \label{eq:t3a}
  \Sigma(K) &=& G_0^{-1}(K)-G^{-1}(K) \nonumber\\
  &=& \sum_Q t(Q)\,G_0(Q-K)\,\varphi^2_{{\bf k}-{\bf q}/2} \;,
\end{eqnarray}
and the T-matrix 
\begin{equation}
  \label{eq:t3b}
  g \;=\; [1+g\,\chi(Q)]\,t(Q)\;,
\end{equation}
where
\begin{equation}
  \label{eq:t3c}
  \chi(Q) \;=\; \sum_K G(K)\,G_0(Q-K)\,\varphi^2_{{\bf k}-{\bf q}/2}
\end{equation}
is the pair susceptibility. Here, $G_0(K) = 1/(i\omega-\epsilon_{\bf k})$ is
the bare propagator, and $K\equiv({\mathbf k},i\omega)$, $\sum_K\equiv
T\sum_{{\mathbf k},i\omega}$, etc., with $\Omega/\omega$ denoting the even/odd
Matsubara frequencies.
Together with the particle number equation

\begin{equation}
  \label{eq:n}
  n \;=\; 2\sum_K G(K)\;,
\end{equation}
\end{mathletters}
Eqs.~(\ref{eq:t3}) form a complete set which, for a given $g$ and $T$,
need to be solved self-consistently for $\Sigma(K)$, $t(Q)$ and the chemical
potential $\mu$.
Equations~(\ref{eq:t3}) represent the basis of the present theory and can be
regarded as a generalized BCS theory in which pair correlations are
explicitly taken into account in the normal state ($T>T_c$) in
a self-consistent fashion.

This approach has implications for the superconducting state as well.  The
presence of (noncritical) pair fluctuations in the supeconducting state
($T<T_c$) leads to progressively stronger deviations from BCS theory as the
coupling is increased.  However, this standard (BCS) theory is embedded in
Eqs.~(\ref{eq:t3}) when the T-matrix is given by

\begin{equation}
  \label{eq:t-sc}
  t_{\text{sc}}(Q) \;=\; \left\{
      \begin{array}{cl}
        0 &\qquad \text{for}\quad T>T_c \,, \\
        -\frac{|\Delta_{\text{sc}}|^2}{T}\delta(Q) & \qquad \text{for}\quad
        T<T_c \,,
      \end{array}
      \right . 
\end{equation}
where $\Delta_{\text{sc}}$ is the superconducting order parameter. Inserting
Eq.~(\ref{eq:t-sc}) into Eqs.~(\ref{eq:t3}) one obtains (i) the usual BCS
self-energy $\Sigma_{\text{BCS}}(K) = |\Delta_{\text{sc}}|^2\varphi^2_{\bf
  k}/(i\omega+\epsilon_{\bf -k})$ and (ii) gap equation $0 =
1+g\chi(0)=1+g\sum_K |\Delta_{\text{sc}}|^2\varphi^2_{\bf
  k}/(\omega^2+E^2_{\bf k})$, where $E_{\bf k}=\sqrt{\epsilon^2_{\bf k} +
|\Delta_{\text{sc}}|^2\varphi^2_{\bf k}}$ is the energy dispersion
of the quasiparticles.  The delta function in $t_{\text{sc}}(Q)$ leads to the
usual (Gor'kov) factorization of the correlation function $C_2(K,K') =
F(K)F(K')$, where the anomalous Green's function
$F(K)=\Delta_{sc}\varphi_{\bf k}G_0(-K)G(K)$.

Under more general circumstances, when pair fluctuations cannot be
neglected, the self-consistent T-matrix in Eqs.~(\ref{eq:t3}) can be written
as

\begin{mathletters}
\label{eq:t4}
\begin{eqnarray}
  \label{eq:t4a}
  t(Q) &=& t_{\text{sc}}(Q) + t_{\text{pg}}(Q) \;,\\
  \label{eq:t4b}
  t_{\text{pg}}(Q) &=& \frac{g}{1+g\chi(Q)}\;,
\end{eqnarray}
\end{mathletters}
where $t_{\text{pg}}(Q)$ is the ``regular'' (or pseudogap) contribution 
[cf.~Eq.~(\ref{eq:t3b})], which should be associated with
(noncritical) pair fluctuations which persist both above and below $T_c$.
At the high temperatures of the normal state, $t_{\text{pg}}(Q)$ is
finite at all $Q$. As the temperature is lowered $t_{\text{pg}}(Q)$
develops a resonant structure corresponding to metastable or long lived
pairs.  Precisely at $T_c$, this quantity becomes divergent for $Q=0$,
in accord with the Thouless criterion for a superconducting pairing
instability.  Once the temperature is less than $T_c$, a nonzero
superconducting order parameter $\Delta_{\text{sc}}$ is established
which obeys the (gap) equation $1+g\chi(0;T)=0$.  It is important to
stress that the same critical temperature is obtained either when
approached from the normal state (using the Thouless criterion)
\begin{equation}
  \label{eq:t5}
  t_{\text{pg}}^{-1}(0) = g^{-1} + \chi(0;T_c) = 0\;,
\end{equation}
or when approached within the superconducting state by setting
$\Delta_{\text{sc}}=0$ within the gap equation.

The calculation of $T_c$ can be substantially simplified by noting that at
this temperature the self-energy is well approximated by \cite{Maly2}

\begin{mathletters}
\label{eq:Sigma-Tca}
\begin{equation}
  \label{eq:Sigma-Tcb}
  \Sigma(K) \approx G_0(-K)\varphi_{\bf k}^2\sum_Q t_{\text{pg}}(Q) \;.
\end{equation}
This approximation is \textit{highly nontrivial} and establishing its
validity requires detailed numerical calculations which are presented
elsewhere.\cite{Maly2}  It should be stressed that Eq.~(\ref{eq:Sigma-Tcb})
cannot be written down on analytical grounds alone.  It is a consequence of
a numerical iterative solution\cite{Maly2} of Eqs.~(\ref{eq:t3}) and
obtains when the T-matrix contains a divergence. In the pseudogap phase
Eq.~(\ref{eq:Sigma-Tcb}) is no longer a valid approximation to
Eq.~(\ref{eq:t3a}).
Indeed, one may arrive at some apparent inconsistencies if this
approximation is used above $T_c$. It follows from
Eq.~(\ref{eq:Sigma-Tcb}) that at $T_c$, the self-energy has the same
symmetry as in the superconducting state. However, it can be seen from
the exact Eq.~(\ref{eq:t3a}), that the pairing symmetry factor
$\varphi_{\bf k}$ cannot, in general, be extracted from inside the
integral. When the integration is properly performed, the anisotropy at
higher temperature will generally be different from that at $T_c$.  This
important consequence, which may be relevant to
experiment,\cite{photoemission} would not follow if
Eq.~(\ref{eq:Sigma-Tcb}) were incorrectly extended beyond its regime of
validity.

The self-energy can be rewritten as
\begin{equation}
 \label{eq:Sigma-Tc}
   \Sigma(K) \approx \frac{\Delta_{\text{pg}}^2\varphi_{\bf k}^2}{i\omega +
    \epsilon_{\bf -k}}\;,
\end{equation}
where the pseudogap parameter is defined as
\end{mathletters}

\begin{mathletters}
\label{eq:SC-set}
\begin{equation}
  \label{Delta-def}
  \Delta^2_{\text{pg}} \equiv -\sum_Q t_{\text{pg}}(Q) = -\sum_{\bf
    q}\int_{-\infty}^{\infty}\frac{d\Omega}{\pi}b(\Omega)\text{Im}
  t_{{\bf q},\Omega}
\end{equation}
in terms of the Bose function $b(\Omega)$. Here, and in what follows, we use
the notation $\Delta_{\text{pg}}$ to represent the amplitude of the pseudogap
at $T_c$, while the momentum dependence is given by
$\Delta_{\mathbf{k}}=\Delta_{\text{pg}} \varphi_{\mathbf{k}}$.
The simple BCS-like form of the self-energy (\ref{eq:Sigma-Tc}) allows us to
express Eqs.~(\ref{eq:t5}) and (\ref{eq:n}) as

\begin{equation}
  t^{-1}_{\text{pg}}(0) = g^{-1} + \sum_{\bf k}\frac{1-2f(E_{\bf
      k})}{2E_{\bf k}}\,\varphi^2_{\bf k} = 0
\label{Thouless-crit}
\end{equation}
and 

\begin{equation}
  n = 2\sum_{\bf k}\left[v^2_{\bf k} + 
    \frac{\epsilon_{\bf k}}{E_{\bf k}}\,f(E_{\bf k})\right] \;,
  \label{number}
\end{equation}
\end{mathletters}
where 
$v^2_{\bf~k}=\frac{1}{2}(1-\epsilon_{\bf~k}/E_{\bf~k}),
u^2_{\bf~k}=\frac{1}{2}(1+\epsilon_{\bf k}/E_{\bf k})$, and $f(E)$ is the
Fermi function.
The complete set of Eqs.~(\ref{eq:SC-set}), which were previously
established in a slightly different form in Ref.~\onlinecite{Maly}, must be
solved self consistently in order to obtain $T_c$, $\mu$, and
$\Delta_{\text{pg}}$ as a function of $g$ and $n$.

The imaginary component of the T-matrix which appears in the pseudogap
equation (\ref{Delta-def}), can be directly calculated by inserting
Eq.~(\ref{eq:Sigma-Tc}) into Eqs.~(\ref{eq:t4b}) and (\ref{eq:t3c}).  However,
extensive numerical calculations\cite{Maly2,JiriThesis} show that as $T_c$
is approached from above, one can simplify this procedure considerably and,
at the same time, obtain considerable physical insight.  For the purposes of
calculating quantities such as $\Delta_{\text{pg}}^2$ (which involve
integrals over the T-matrix), it suffices to approximate the T-matrix near
$T_c$ by its values at small $\Omega$ and $\mathbf{q}$.  We derive the
appropriate form by noting that the inverse of the analytically continued
function can be written as
\begin{eqnarray}
  \text{Re}\, t^{-1}_{{\mathbf{q}},\Omega} &\approx& a'_0\,(\Omega-
  \Omega_{\mathbf{q}}) \;, \nonumber\\
  \text{Im}\, t^{-1}_{{\mathbf{q}},\Omega} &\approx& a''_0\,\Omega 
\;, 
\nonumber
\end{eqnarray}
where $\Omega_{\mathbf{q}}$ can be naturally interpreted as an energy
dispersion of pairs of fermions. Here the parameters $a'_0$ and $a''_0$ are
essentially constant in the relevant range of momentum and frequency.
Moreover,\cite{Maly2} independent of the values of $g$, sufficiently close
to $T_c$, we find that the ratio $\varepsilon \equiv a''_0/a'_0 \ll 1$.
Hence the imaginary part of the T-matrix can be approximated as

\begin{eqnarray}
  \label{eq:Im-t}
  \text{Im}\, t_{{\mathbf{q}},\Omega} &\approx& - 
\lim_{\varepsilon\rightarrow 0}
  \frac{\text{Im}\, t^{-1}_{{\mathbf{q}},\Omega}}{
    \left[\text{Re}\, t^{-1}_{{\mathbf{q}},\Omega}\right]^2 +
  \left[\text{Im}\, t^{-1}_{{\mathbf{q}},\Omega}\right]^2 } \nonumber 
\\
  &=& - \frac{1}{a'_0\,\Omega} \lim_{\varepsilon\rightarrow 0}
  \frac{\varepsilon}{\left(1-\Omega_{\mathbf{q}}/\Omega \right)^2 +
  \varepsilon^2} \nonumber \\
  &=& -\frac{\pi}{a'_0}\,\delta(\Omega-\Omega_{\mathbf{q}})\;, 
\nonumber
\end{eqnarray}

Finally, it is useful to rewrite the approximated T-matrix at $T_c$
(for small momenta and frequencies) in a more compact form as

\begin{equation}
  \label{eq:T-model}
  t_{\mathbf{q},\Omega} 
  \approx \frac{a_0^{\prime -1}}{\Omega-\Omega_{\mathbf{q}}
        +i\epsilon\Omega}\;,
\end{equation}
This approximated T-matrix takes the natural form of a pair ``Green's
function" or propagator, with characteristic dispersion
$\Omega_{\mathbf{q}}$. Using the inversion symmetry of the Hamiltonian
we deduce that $\Omega_{\mathbf{q}}$ varies quadratically at $T_c$ with
the wave vector, as
\begin{equation}\large
\Omega_{\mathbf{q}}\approx \left\{
\begin{array}{l@{\quad}l}
\frac{q^2}{2M^*} & \text{\normalsize for 3d},\\
&\\
\frac{q_\parallel^2}{2M^*_\parallel}+\frac{q_\perp^2}{2M^*_\perp}
              &  \text{\normalsize for quasi-2d},
\end{array}
\right.
\label{Bq2}
\end{equation}
where $M^*$ is the effective mass of the pairs. By quasi-2d we mean a
highly anisotropic 3d system with $M_{\perp}^*/M_{\parallel}^*\gg 1$.
The effective pair mass is determined by an expansion of the pair
dispersion, given by

\begin{eqnarray}
\label{mass}
\Omega_{\mathbf{q}}& =& -\frac{1}{a_0^\prime}\left\{
        \sum_{\mathbf{k}}\left[\frac{1-f(E_{\mathbf{k}})-
f(\epsilon_{\mathbf{
        k-q}})}{E_{\mathbf{k}}+\epsilon_{\mathbf{k-
q}}}u_{\mathbf{k}}^2
        \right. \right. \\
        &&\left. \left.
        -\frac{f(E_{\mathbf{k}})-f(\epsilon_{\mathbf{
        k-q}})}{E_{\mathbf{k}}-\epsilon_{\mathbf{k-
q}}}v_{\mathbf{k}}^2\right]
        \varphi^2_{{\mathbf k-q}/2}
        -\frac{1-
2f(E_{\mathbf{k}})}{2E_{\mathbf{k}}}\varphi^2_{\mathbf{k}}
        \right\}\;,\nonumber
\end{eqnarray}
where 
\begin{equation}
a_0^\prime=\frac{1}{2\Delta_{\text{pg}}^2}\sum_{\mathbf{k}}
        \left[\left[1-2f(\epsilon_{\mathbf{k}})\right]
        -\frac{\epsilon_{\mathbf{k}}}{E_{\mathbf{k}}}\left[
        1-2f(E_{\mathbf{k}})\right]\right] \;.
\end{equation}
We have verified numerically that the above leading order (in ${\bf q}$)
contributions in Eq.~(\ref{Bq2}) dominate in our subsequent calculations, so
that higher order terms in the expansion can be dropped.  The relationship
between the effective mass of the pairs and $T_c$ will be explored in detail
in Sec.~\ref{subsec:overview}.

The above analysis will be applied to isotropic and anisotropic jellium with
$s$-wave pairing, as well as to discrete lattices. In the latter case we
consider both $s$- and $d$-wave symmetry of the pairing interaction
$V_{\mathbf{k,k'}}$.  The distinction between these various situations
enters via the dispersion relation $\epsilon_{\mathbf{k}}$ and the symmetry
factor $\varphi_{\mathbf{k}}$, which will be characterized below, according
to the details of the physical system. For definiteness, in our quasi-2d
calculations it is assumed that the pairing interaction depends only on the
in-plane momenta.

(i) {\em 3d jellium, s-wave symmetry\/}.  
We assume a parabolic dispersion relation,
$\epsilon_{\mathbf{k}}={\mathbf{k}}^2/2m-\mu$, with $\varphi_{\mathbf{k}} =
(1+k^2/k_0^2)^{-1/2}$.  The parameter $k_0$ is the inverse range of the
interaction and represents a soft cutoff in momentum space for the
interaction. As will be clear later, $k_0 > k_F$ is assumed in general in
order to access the strong coupling limit.  It is convenient to introduce a
dimensionless scale $g/g_c$ for the coupling constant. Here, following
Ref.~\onlinecite{NSR}, we choose $g_c = -4\pi/mk_0$, which corresponds to
the critical value of the coupling above which bound pairs are formed in
vacuum.

(ii) {\em 2d jellium, s-wave symmetry}. For 2d jellium we choose the
same $\epsilon_{\mathbf{k}}$ and $\varphi_{\mathbf{k}}$ as for case (i).
We find that $T_c=0$, in agreement with the Mermin-Wagner theorem. To
understand this result, note that the assumption that $T_c$ is finite
leads to a contradiction, associated with an unphysical divergence in
the pseudogap amplitude. This unphysical result derives from an
infrared, logarithmic divergence in the phase space integral on the
right hand side of Eq.~(\ref{Delta-def}). This divergence can be made
obvious by rewriting this equation using the low frequency, long
wavelength expansion of the T-matrix so that
\begin{equation}
  \label{eq:PG-model}
  \Delta_{\text{pg}}^2 \approx \frac{1}{a'_0}\sum_{\mathbf{q}}
  b(\Omega_{\mathbf{q}})\;. 
\end{equation}
Pairing fluctuations, thus, disorder the system for any finite
temperature.  Even in 2d, for which $T_c =0 $, we obtain a finite
pseudogap, as will be seen in Sec.~\ref{subsec:dimensionality}. It
should be noted that this result is general and remains valid for both
$s$- and $d$-wave pairing on discrete lattices, as well.  This is a
consequence of the fact that the lattice energy dispersion is quadratic
at sufficiently small wavevectors, so that the same arguments as above
can be applied.

(iii) {\em Quasi-2d jellium, s-wave symmetry}. Here we use
$\varphi_{\mathbf{k}}$, as in the previous two cases, and adopt an
anisotropic energy dispersion

\begin{equation}
  \label{eq:Epsilon-2d-jellium}
  \epsilon_{\mathbf{k}} = 
\frac{{\mathbf{k}}^2_{\parallel}}{2m_\parallel} +
  \frac{k_{\perp}^2}{2m_{\perp}} - \mu\;,
\end{equation}
where $k_\perp$ is restricted to a finite interval
($|k_\perp|\le\pi$),\cite{kz-cutoff} while $\mathbf{k}_{\parallel}$ is
unconstrained.

By tuning the value of the anisotropy ratio $m_{\perp}/m_\parallel$
from one to infinity, this model can be applied to study effects
associated with continuously varying dimensionality from 3d to
2d.\cite{quasi-2d-note} For convenience, we use the parameter $g_c$
derived for 3d jellium, as a scale factor for the coupling strength,
and call it $g_0$ to avoid confusion.

(iv) {\em Quasi-2d lattice, s-  and d-wave symmetry}.  In the 
presence of a lattice we will adopt  a simple tight-binding model  
with dispersion 
\begin{equation}
  \label{eq:Epsilon-2d-lattice}
  \epsilon_{\mathbf{k}} = 2\,t_\parallel (2-\cos{k_x}-\cos{k_y}) +
  2\;t_{\perp}(1-\cos{k_{\perp}}) - \mu\;,
\end{equation}
where $t_\parallel$ ($t_{\perp}$) is the hopping integral for the
in-plane (out-of-plane) motion.  Here we consider both isotropic
$s$-wave pairing symmetry with $\varphi_{\mathbf{k}}=1$, as in the
negative U Hubbard model, as well as $d$-wave effects with

\begin{equation}
  \label{eq:d-wave}
  \varphi_{\mathbf{k}} = \cos{k_x}-\cos{k_y}\;.
\end{equation}
It should be noted that in the lattice case, because the momentum
integration is restricted to the first Brillouin zone, it is not
necessary to introduce a cutoff for the interaction in momentum space.

\section{Numerical Results}
\label{sec:Results}

Equations~(\ref{eq:SC-set}), together with the various models for
$\varphi_{\mathbf{k}} $ and $\epsilon_{\mathbf{k}}, $ were solved
numerically for $\Delta_{\text{pg}}$, $\mu$, and $T_c$. The numerically
obtained solutions satisfy the appropriate equations with an accuracy
higher than $10^{-7}$.  The momentum summations were calculated through
numerical integration over the whole ${\bf k}$ space for the jellium
case, and over the entire Brillouin zone for the lattice. However, to
facilitate our calculations in the case of the quasi-2d lattice with a
$d$-wave pairing interaction, the momentum integral along the
out-of-plane direction was generally replaced by summation on a
lattice with $N_\perp=16$ sites. For completeness we compared solutions
obtained with and without the low frequency, long wavelength expansions
of the T-matrix discussed above, and found extremely good agreement
between the two different approaches.  In general, we chose the ratios
$m_\perp/m_\parallel =100$ or $t_\perp/t_\parallel=0.01$, although
higher values of the anisotropy
were used for illustrative purposes in some cases.%

\subsection{Overview:  $T_c$ and effective mass of the pairs}
\label{subsec:overview}

In was pointed out in Ref.~\onlinecite{NSR} in the context of the
attractive Hubbard calculations, that the appropriate description of
the strong coupling limit corresponds to \textit{interacting} bosons
on a lattice with effective hopping integral $t' \approx
-2t^2/U$. It, therefore, will necessarily vanish in the strong
coupling limit, as $U \rightarrow \infty$.  In addition to this
hopping, there is an effective boson-boson repulsion which also varies
as $V^\prime \approx -2t^2/U$.

This description of a boson Hamiltonian can be related to the present
calculations through Eqs.~(\ref{eq:T-model}) - (\ref{mass}) which
represent the Green's function for such a Hamiltonian and its
parametrization via the pair mass $M^*$. By solving
Eqs.~(\ref{eq:SC-set}) self-consistently and identifying $M^*$ from the
effective pair propagator (or T-matrix), our $M^*$ necessarily
incorporates all renormalizations such as Pauli principle induced
pair-pair repulsion, pairing symmetry and density related effects. Note,
in contrast to Ref.~\onlinecite{NSR}, in the present work we are not
restricted to the bosonic limit, nor is it essential to consider a
periodic lattice. Thus, much of this language is also relevant to the
moderately strong coupling (but still fermionic) regime, and can even be
applied to jellium.

The goal of this subsection is to establish a natural framework for
relating $M^*$ to $T_c$.  The parameters which enter into $M^*$ via
Eq.~(\ref{mass}) vary according to the length scales in the various
physical models. In the case of jellium, $M^*$ depends in an important
way on the ratio $k_0/k_F$.  For the case of $s$-wave pairing on a
lattice, $M^*$ depends on the inverse lattice constant $\pi/a$ and
density $n$.  Finally, for the case of $d$-wave pairing, there is an
additional length scale introduced as a result of the finite spatial
extent of the pair. This enters as if there were an equivalent
reduction in $k_0/k_F$ in the analogous jellium model.  The following
factors act to increase $M^*$ or, alternatively, to reduce the
mobility of the pairs: the presence of a periodic lattice, a spatially
extended pairing symmetry (such as $d$-wave) or, for jellium, small
values of the ratio $k_0/k_F \lesssim 0.4$ (i.e., high density).

In order to relate $T_c$ to $M^*$, we observe that in an ideal
Bose-Einstein system $T_c$ is inversely proportional to the mass.
Here, this dependence is maintained, in a much more complex theory, as
a consequence of Eqs.~(\ref{Delta-def}) and (\ref{eq:PG-model}). This
is essentially an equation for the number of pairs (bosons), with
renormalized mass $M^*$.  Thus, as we increase $g$ towards the bosonic
regime, it is not surprising that $T_c$ varies inversely with $M^*$.

This leads to our main observations, which apply to moderate and large
$g$, although not necessarily in the strict bosonic regime.  (i) For the
general lattice case, we find that $T_c$ vanishes, either asymptotically
or abruptly, as the coupling increases, in the same way that the inverse
pair mass approaches zero.\cite{zeroTc} (ii) For the case of jellium or
low densities on a lattice, both $T_c$ and $M^*$ remain finite and are
inversely proportional. These observations are consistent with, but go
beyond, the physical picture in Ref.~\onlinecite{NSR} that $T_c$ is
expected to be proportional to the pair hopping integral $t'$.  It
should be stressed that in the very weak coupling limit the pair size or
correlation length is large. In this case, the motion of the pairs
becomes highly collective, so that the effective pair mass is very
small.

In the presence of a lattice, the dependence on band filling $n$ is
also important for $M^*$, and thereby, for $T_c$. We find that the bosonic
regime is not accessed for large $n > n_c\approx 0.53$. There are
two reasons why superconductivity abruptly disappears within the
fermionic regime. This occurs primarily (in the language of
Ref.~\onlinecite{NSR}) as a consequence of large pair-pair repulsion,
relevant for high electronic densities, which leads to large $M^*$. In
addition, there are effects associated with the particle-hole symmetry
at half filling.\cite{CDW} Precisely at half filling (i.e., the
``filling factor'' $f = 1/2$, or $ 2f= n = 1$), for the band structure
we consider, there is complete particle-hole symmetry and $\mu$ is
pinned at $E_F$.  Similarly, in the vicinity of $n = 1 $, the chemical
potential remains near $E_F$ for very large coupling constants $g$.

By contrast, in the small density (lattice) limit for the $s$-wave
case, ($n\approx 0.1$), pair-pair repulsion is relatively unimportant
in $M^*$ and there is no particle-hole symmetry. In this way the
bosonic regime is readily accessed. Moreover, in this limit we see a
precise scaling of $T_c$ with $1/g$ in the same way as predicted by
Ref.~\onlinecite{NSR} (via the parameter $t' = -2t^2/U$).  Thus in
this low density limit superconductivity disappears asymptotically,
rather than abruptly.

The effects of pairing symmetry should also be stressed. Because of the
spatial extent of the $d$-wave function, the pair mobility is strongly
suppressed, and, thus, $M^*$ is relatively larger than for the $s$-wave
case. This lower mobility of $d$-wave pairs leads to the important
result that superconductivity is \textit{always abruptly} (rather than
asymptotically) destroyed with sufficiently large coupling.  Near half
filling we find $\mu$ remains large when $T_c$ vanishes, at large $g$.
As the density $n$ is reduced, \linebreak

\begin{figure}[tbh]
\vskip -0.4in
\centerline{\leavevmode
\epsfxsize=3.3in
\epsfbox{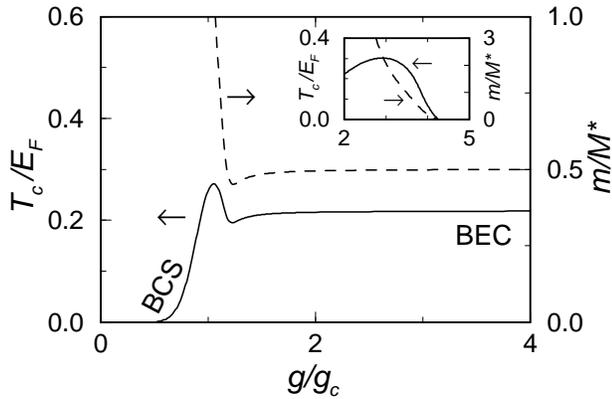}
}
\caption{$T_c$ and $m/M^*$ as a function of $g/g_c$ in the 3d 
jellium model with $k_0/k_F=4$ (main figure) and 
$k_0/k_F=1/3$ (inset), corresponding to short range  (or low 
density), and long range interactions (or high density), respectively. 
}
\label{Fig1}
\end{figure}

\noindent
away from half filling, $\mu$ decreases
somewhat. It is important to note that the system remains in the
fermionic regime (with positive $\mu$) for all densities down to
$n\approx 0.09$.

In all cases discussed thus far, $T_c$ exhibits a nonmonotonic
dependence on the coupling constant. It grows exponentially at small $g$
and shuts off either asymptotically or abruptly at higher $g$.  One can
view this effect as deriving from a competition between pairing energy
scales and effective mass or mobility energy scales.  This competition
is not entirely dissimilar to that found in more conventional Eliashberg
theory where the fermionic renormalized mass and the attractive
interaction compete in such a way as to lead to a saturation in $T_c$ at
large coupling.  However, in the present context, for intermediate and
strong coupling, we are far from the Fermi liquid regime and the
effective mass of the quasibound or bound pair is a more appropriate
variable.

With this background, it should not be surprising that nonmonotonic
behavior will arise, even in situations as simple as in jellium
models.  Indeed, in this case we find that for sufficiently long range
interactions or high densities (small $k_0/k_F$) superconductivity
disappears abruptly before the bosonic regime can be
reached.\cite{Itmaybereentrant} Even for the case of short range
interactions ($k_0/k_F = 4$), there is a depression in $T_c$ caused
by an increase in the pair mass, while still in the fermionic regime.

In Fig.~\ref{Fig1} we plot the calculated $T_c$ for the case of an
isotropic, 3d jellium model with $s$-wave pairing, along with the inverse
pair mass $m/M^*$.  This figure is presented primarily as a base line with
which to compare subsequent plots. The parameter $k_0/ k_F = 4$, is
reasonably large so that the high $g$ asymptote is found to reach the ideal
Bose-Einstein limit ($T_c=0.218E_F$) with $M^*=2m$.  The approach to the
high $g$ asymptote is from below, as is expected.\cite{Haussmann} This is a
result of the decreasing Pauli principle repulsion associated with
increasing $g$, and concomitant reduction in pair size. The nonmonotonic
behavior at intermediate $g/g_c \approx 1 $ can be associated with structure
in the effective pair mass, and has been discussed previously from a
different perspective.\cite{Maly2}

In the inset are plotted analogous curves for the case of long range
interactions or high densities ($k_0/k_F = 1/3$).  This figure
illustrates how superconductivity vanishes abruptly before the bosonic
regime is reached, as a consequence of a diverging pair
mass.\cite{Itmaybereentrant} \vspace*{-0.01in}

\subsection{Effects of dimensionality}
\label{subsec:dimensionality}

In this subsection we illustrate the effects of anisotropy or
dimensionality on $T_c$ (and on $\Delta_{\text{pg}}$ and $\mu$ ) within
the context of a jellium dispersion.\cite{jellium+lattice} A
particularly important check on our theoretical interpolation scheme
is to ascertain that $T_c$ is zero in the strict 2d limit and that
$\mu$ varies continuously from $E_F$ in weak coupling to the large
negative values characteristic of the strong coupling bosonic limit.
The present calculational scheme should be compared with that of
Yamada and co-workers\cite{Yamada} who included ``mode coupling" or
feedback contributions to $T_c$, but only at the level of the lowest
order ``box'' diagram discussed in Ref. \onlinecite{Janko}. These
authors were unable to find a smooth interpolation between weak and
strong coupling, but did successfully repair the
problems\cite{RVS,SereneNSR} associated with the NSR scheme, which led to
negative $\mu$ even in arbitrarily weak coupling.

\begin{figure}
\vspace*{-0.2in}
\centerline{\leavevmode
\epsfxsize=3.2in
\epsfbox{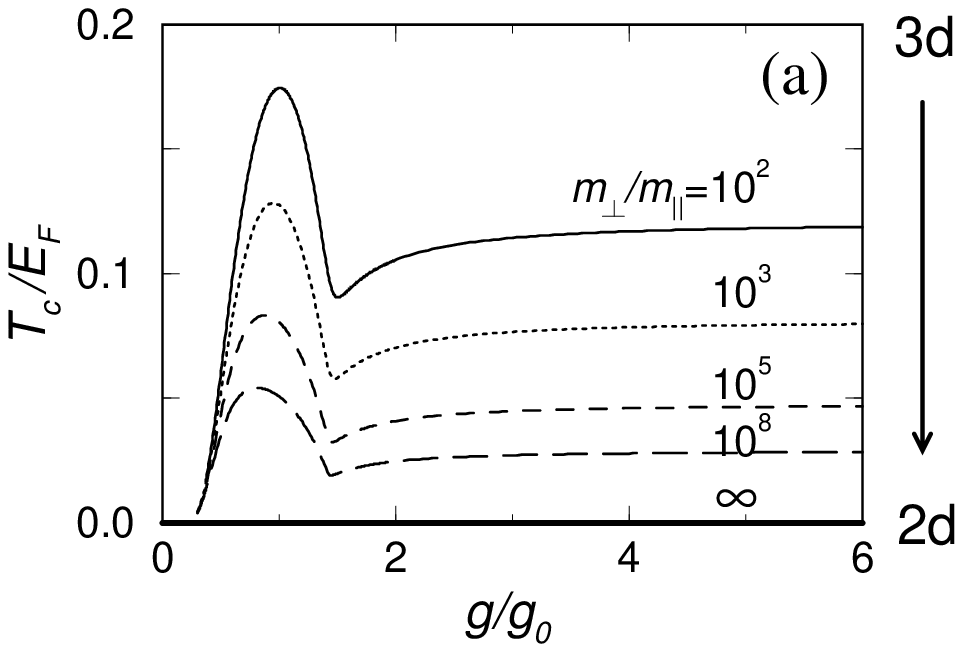}
}
\centerline{\leavevmode\hskip -0.57cm
\epsfxsize=3.in
\epsfbox{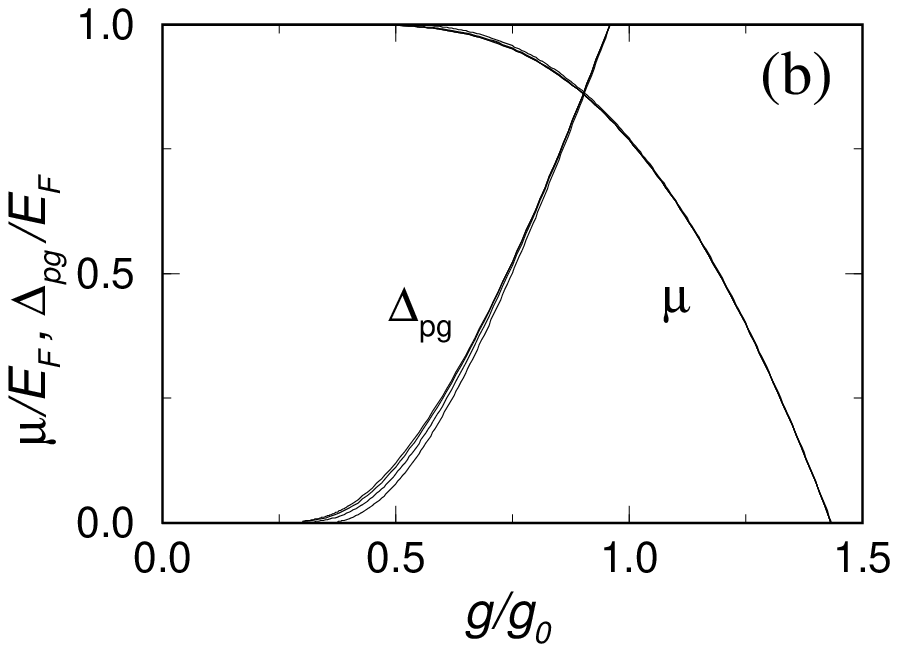}
}
\centerline{\leavevmode\hskip -0.56cm
\epsfxsize=3.05in
\epsfbox{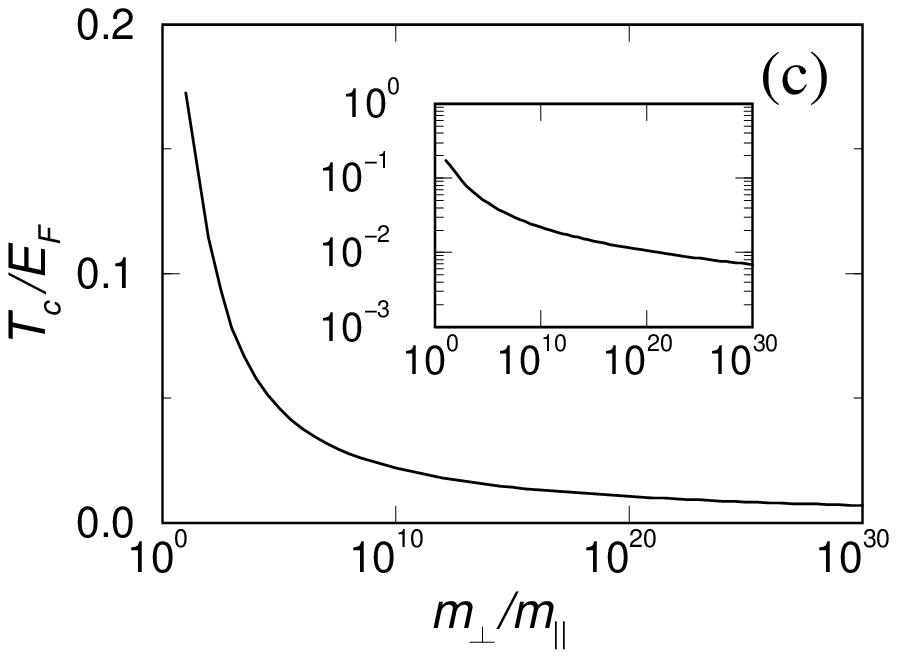}
}
\caption{Dimensionality crossover in a quasi-2d jellium model.  (a)
$T_c$ as a function of $g$ is seen to vanish for all $g$ as
$m_\perp/m_\parallel \rightarrow \infty$, while (b) $\mu$ and
$\Delta_{\text{pg}}$ change little.
A continuous variation of $T_c$
versus $m_\perp/m_\parallel$ at $g/g_0=4$ is shown in the main portion
(semi-log plot) and the inset (log-log plot) of (c).  Here $k_0/k_F=4,
g_0\equiv -4\pi/mk_0$.  }
\label{Fig2}
\end{figure}

Figures \ref{Fig2}(a) and \ref{Fig2}(b) show the effect on $T_c$ and on
$\Delta_{\text{pg}}$ and $\mu$, respectively, of introducing a layering or
anisotropy into jellium with $s$-wave pairing. The various curves correspond
to different values of the anisotropy ratio $m_\perp/m_\parallel$. It can be
seen from these two figures that $T_c$ approaches zero as the dimensionality
approaches 2.  At the same time the chemical potential $\mu$ interpolates
smoothly from the Fermi energy at weak coupling towards zero at around
$g/g_0 = 1.5$ to large negative values (not shown) at even larger $g$.  The
vanishing of the superconducting transition in strictly 2d was discussed in
detail in Sec.~\ref{sec:Theory}.

It should be noted that quasi-two dimensionality will be an important
feature as we begin to incorporate the complexity of $d$-wave pairing.
The essential physics introduced by decreasing the dimensionality is
the reduction in energy scales for $T_c$. The chemical potential and
pseudogap amplitude are relatively unaffected by dimensional
crossover effects.\cite{pseudo-note} While $T_c$ rapidly falls off
when anisotropy is first introduced into a 3d system (such as is
plotted in Fig.~\ref{Fig1}), the approach to the strict 2d limit is
logarithmic and therefore slow, as can be seen explicitly in
Fig.~\ref{Fig2}(c). Thus, in this regime, to get further significant
reductions in $T_c$ associated with a dimensionality reduction
requires extremely large changes in the mass anisotropy.

\subsection{Effects of a periodic lattice}

The first applications of a BCS Bose-Einstein crossover theory to a
periodic lattice were presented in Ref.~\onlinecite{NSR}. The present
approach represents an extension of the NSR theory in two important
ways: we introduce mode coupling or full self-energy effects which are
parametrized by $\Delta_{\text{pg}}$, and which enter via
Eq.~(\ref{Delta-def}).  Moreover, the number equation [see
Eq.~(\ref{number}),
\linebreak

\noindent
 which  is a rewriting of Eq.~(\ref{eq:n}) in terms of
$\Delta_{\text{pg}}$] is evaluated by including self-energy effects to
all orders.  This is in contrast  to the approximate number equation
used in Ref.~\onlinecite{NSR}, which
 includes only the first order
correction.  In this way we are able to capture the effects which were
qualitatively treated by these authors and which are associated with
the lattice.

\begin{figure}
\vspace*{-0.3in}
\centerline{\leavevmode\hskip -0.5cm
\epsfxsize=3.05in
\epsfbox{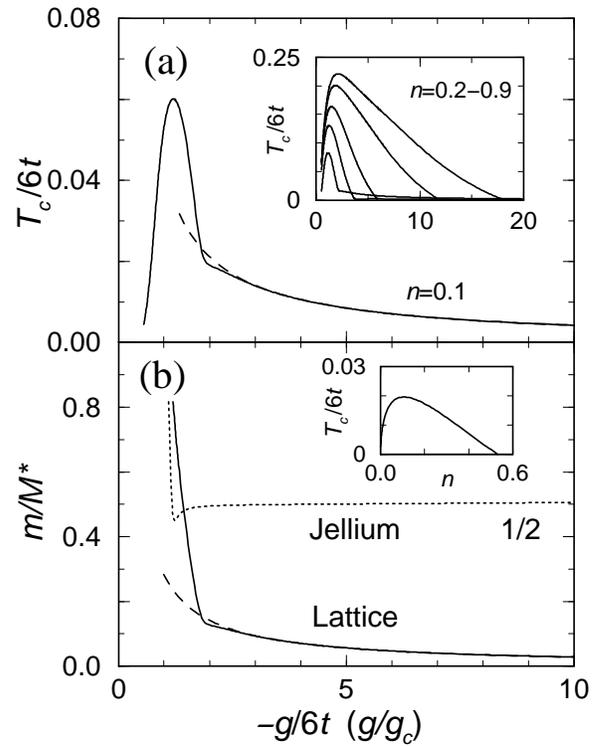}
}
\caption{(a) $T_c$ and (b) $m/M^*$ (solid lines) \textit{vs} $g$ 
at low filling ($n=0.1$) on a 3d lattice, and  $T_c$ at larger filling in
the inset of (a). A fit to the functional form  
$t^\prime=-2t^2/g$ is plotted (dashed lines) in (a) and (b) with
adjusted proportionality constants.
 For comparison, $m/M^*$
vs $g/g_c$ for 3d jellium (Fig.~\ref{Fig1}) is replotted
(dotted line) in (b).  From bottom to top, the inset of (a) shows
$T_c$ for densities $n=0.2, 0.5, 0.7, 0.85$, and 0.9. The inset of
(b) shows $T_c$ at $\mu=0$ as a function of $n$.
}
\label{Fig3}
\end{figure}

Figure \ref{Fig3}(a) plots the behavior of $T_c$ (solid line) in an
isotropic three-dimensional lattice (with $s$-wave pairing,
$\varphi_{\bf k}=1$) at a low density $n=0.1$.  The effects of higher
\linebreak

\noindent
electronic filling are shown in the inset. The low $n$ behavior in the
main portion of the figure can be compared with the jellium calculations
of Fig.~\ref{Fig1}.  For small $n$, $T_c$ decreases asymptotically to
zero at high $g$.  For larger $n$, $T_c$ vanishes abruptly before the
bosonic regime ($\mu < 0$) is reached [See inset of Fig.~\ref{Fig3}(b)].
These various effects reflect the analogous reduction in the effective
pair mobility, parametrized by the inverse pair mass $m/M^*$. To see
the correlation with $m/M^*$ in the low density limit, we plot this
quantity in Fig.~\ref{Fig3}b, for the lattice as well as jellium case
(where for the latter, $m/M^*\rightarrow 1/2$ at large $g$). Here the
coupling constants are indicated in terms of $g/g_c$ for jellium and
$-g/6t$ for the lattice.  The inflection points at $-g/6t\approx 2$ in
both $T_c$ and $m/M^*$ curves correspond to $\mu=0$, which marks the
onset of the bosonic regime.

Also plotted in both Fig.~\ref{Fig3}(a) and \ref{Fig3}(b) (dashed lines)
is the effective hopping $t^\prime=-2t^2/g$ for $n=0.1$, rescaled such
that it coincides with $T_c$ and $m/M^*$, respectively, at high coupling
($-g/6t=30$).  This figure illustrates clearly the effect first noted by
Nozi\`eres and Schmitt-Rink that in the entire bosonic regime, $T_c$
varies with high precision as $t^\prime$ or equivalently as $m/M^*$.

Finally, in the inset of Fig.~\ref{Fig3}(b), we demonstrate the limiting
value of $n$, above which the bosonic limit can not be \linebreak

\begin{figure}
\vskip -0.25in
\centerline{\leavevmode\hskip -0.4cm
\epsfxsize=3.in
\epsfbox{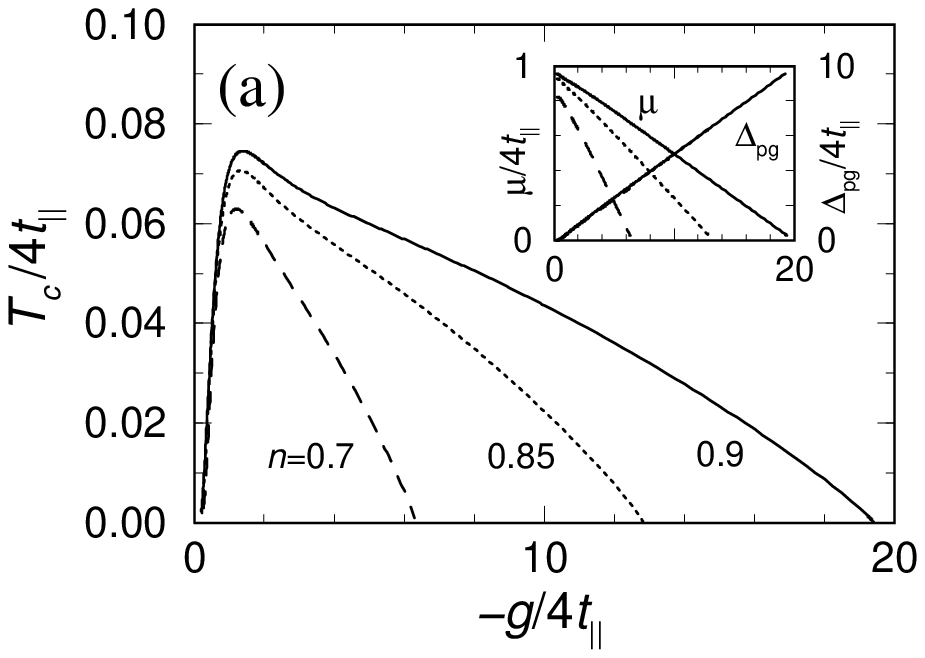}
}
\centerline{\leavevmode\hskip -0.4cm
\epsfxsize=3.in
\epsfbox{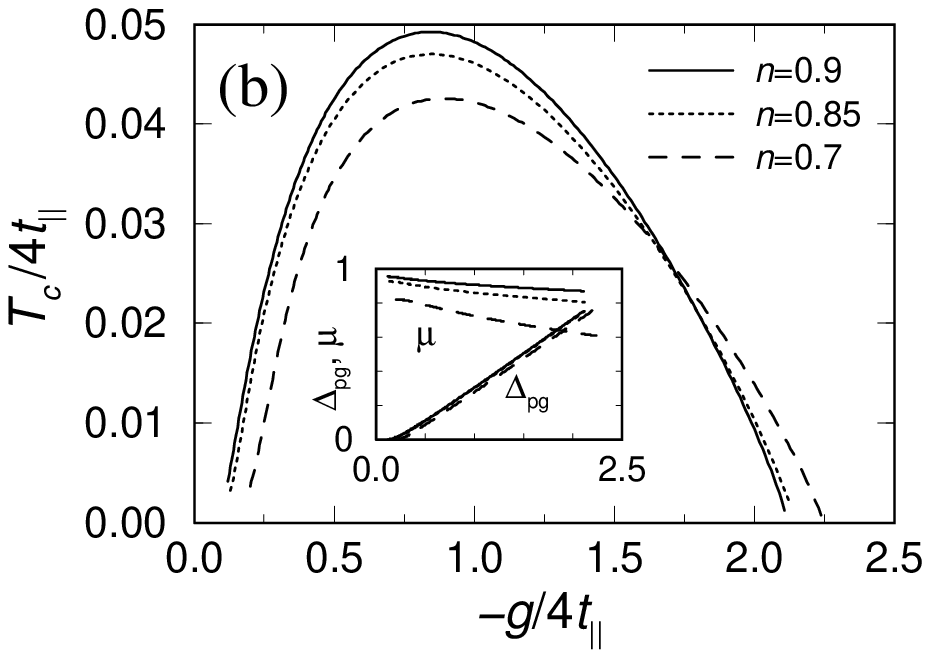}
}
\caption{Lattice effects on $T_c$ (main figure) and $\mu$ and
$\Delta_{\text{pg}}$ (inset) as a function of $g$ for $n$=0.7 (dashed
lines), 0.85 (dotted lines), and 0.9 (solid lines) in quasi-2d for (a)
$s$-wave and (b) $d$-wave pairing symmetries.  Here
$t_\perp/t_\parallel=0.01$. In (b), $T_c$ vanishes at a much smaller
$g$ than does its $s$-wave counterpart.  }
\label{Fig4}
\end{figure}

\noindent
accessed.  What
is plotted here is the value of $T_c$ at which $\mu$ is zero as a
function of density $n$.  This figure indicates that the bosonic regime
can not be reached for $n>n_c\approx 0.53$.  At densities higher than
this, the pair-pair repulsion increases $M^*$ sufficiently, so that
$T_c$ vanishes abruptly, while $\mu$ is still positive.

\subsection{Effects of \textit{d}-wave symmetry}

We now introduce the effects of a $d$-wave pairing interaction. For the
purposes of comparison we begin by illustrating $T_c$ for the case of
$s$-wave pairing on an anisotropic lattice, shown in Fig.~\ref{Fig4}(a), for
three different values (0.7, 0.85, and 0.9) of the density
$n$.\cite{n-choice} The inset indicates the behavior of the pseudogap
magnitude and the chemical potential. The plots of $\Delta_{\text{pg}}$ for
the three different $n$ are essentially unresolvable in the figure. Note,
from the inset, that within small numerical errors $T_c$ and $\mu$ vanish
simultaneously.  A comparison of the magnitude of $T_c$ (in the main figure)
with the 3d counterpart shown in the inset of Fig.~\ref{Fig3}(a) illustrates
how $T_c$ is suppressed by quasi-two dimensionality.\cite{low-n-s}

In Fig. \ref{Fig4}(b), similar plots are presented for the $d$-wave case.
Here we use the same values of the filling factor as in
Fig.~\ref{Fig4}(a), to which Fig.~\ref{Fig4}(b) should be compared. The
essential difference between the two figures is the large $g$ behavior.
Lattice effects produce the expected cutoff for $s$-wave pairing. In
the $d$-wave situation this cutoff is at even smaller $g$, and
moreover, corresponds to $\mu\approx E_F$.  Calculations similar to
those shown in the inset of Fig.~\ref{Fig3}(b) indicate that
superconductivity disappears while $\mu$ remains positive for all $n$
above the extreme low density limit (i.e., for $n > n_c \approx
0.09$).\cite{low-n} This behavior is in contrast to that of the $s$-wave
case where $n_c \approx 0.53$.

In the $d$-wave case, the pair size cannot be made arbitrarily small,
no matter how strong the interaction.  As a result of the extended size
of the pairs, residual repulsive interactions play a more important
role. In this way, the pair mobility is reduced and the pair mass
increased. Thus, as a consequence of the finite pair size, \textit{in
 the d-wave case the system essentially never reaches the
 superconducting bosonic regime}.\vspace*{-0.01in}

\subsection{Phase diagrams}

In this section we introduce an additional energy scale $T^*$, and in
this way, arrive at plots of characteristic ``phase diagrams" for the
crossover problem. Our focus is on the pseudogap onset, so that
attention is restricted to relatively small and intermediate coupling
constants $g$; consequently, the 
bosonic regime is not addressed. Here,
our calculations of $T^*$ are based on the solution of
Eq.~(\ref{Thouless-crit}), along with Eq.~(\ref{number}), under the
assumption that $\Delta_{\text{pg}} = 0$.  This approximation for $T^*$
is consistent with more detailed numerical work\cite{Maly2} in which
this temperature is associated with the onset of a pair resonance in the
T-matrix.

\begin{figure}
\centerline{\leavevmode\hskip -0.4cm
\epsfxsize=2.8in
\epsfbox{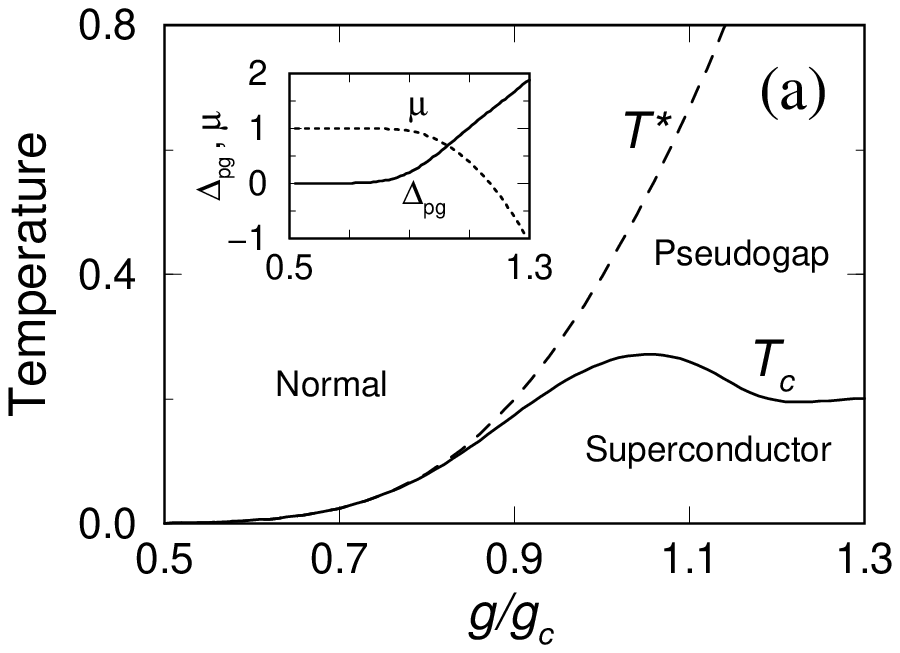}
}
\centerline{\leavevmode\hskip -0.4cm
\epsfxsize=2.8in
\epsfbox{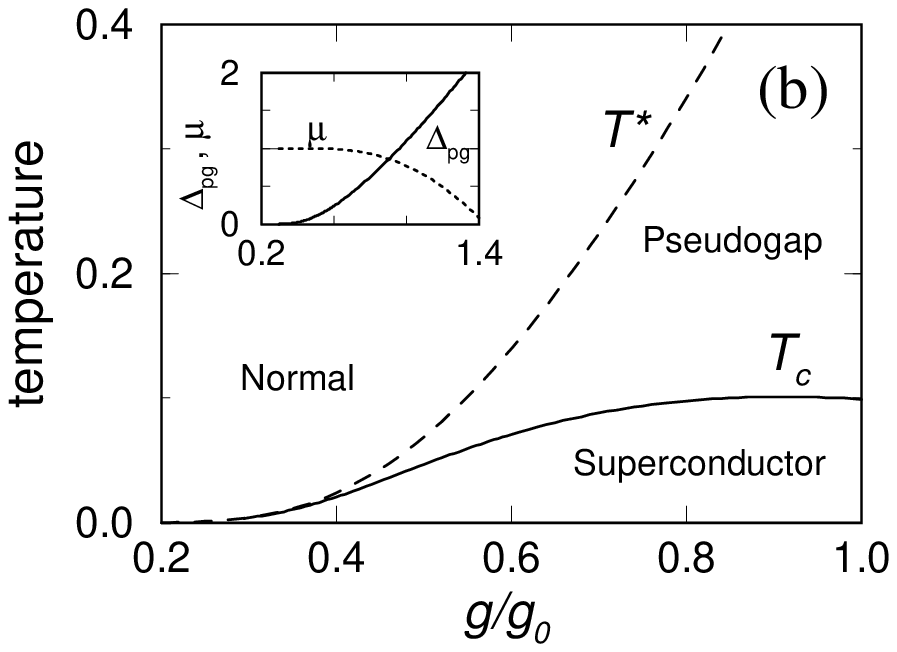}
}
\centerline{\leavevmode\hskip -0.4cm
\epsfxsize=2.8in
\epsfbox{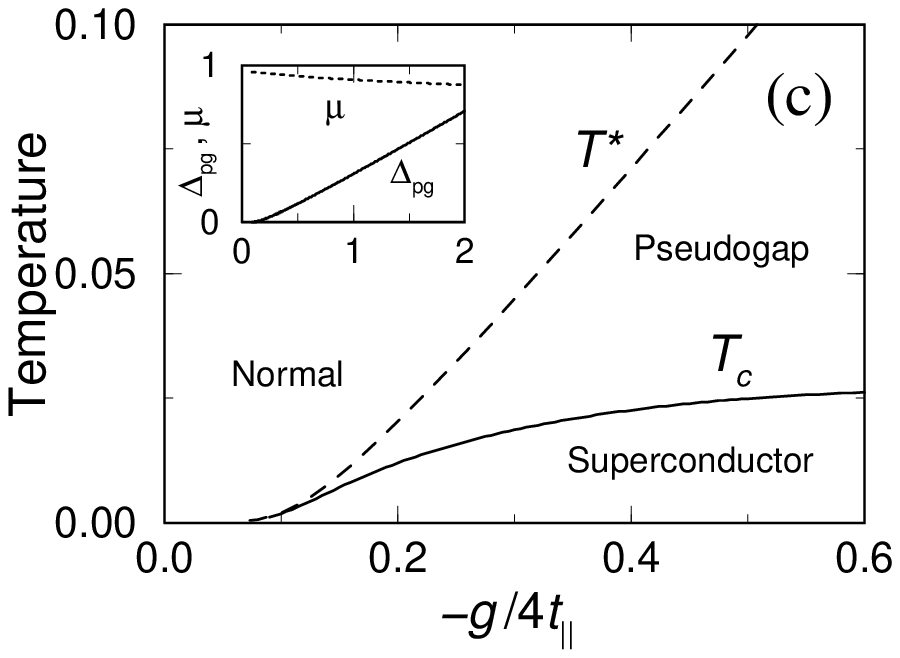}
}
\caption{Phase diagrams of (a) 3d jellium ($k_0/k_F=4$), quasi-2d
jellium ($k_0/k_F=4, m_\perp/m_\parallel=10^4$), (c) a $d$-wave
symmetry on a quasi-2d lattice 
($t_\perp/t_\parallel=10^{-4}$). Here we take $ n=0.9$.
The same energy units are (a), (b) $E_F$ and
(c) $4 t_\parallel$.}
\label{Fig5}
\end{figure}

In Figs.~\ref{Fig5}(a)-\ref{Fig5}(c) our results are consolidated into
phase diagrams for the different physical situations. The case of 3d
jellium, with $s$-wave pairing [Fig.~\ref{Fig5}(a)] is presented
primarily as a point of comparison.  Figure~\ref{Fig5}(b) corresponds to
quasi-2d jellium ($m_\perp/m_\parallel = 10^4$), with $s$-wave pairing
and Fig.~\ref{Fig5}(c) to the case of $d$-wave pairing in a quasi-2d
lattice case ($t_\perp/t_\parallel = 10^{-4}$).\cite{Slatticephase} The
insets indicate the behavior of $\mu$ and $\Delta_{\text{pg}}$.
Comparing $T^*$ with $T_c$ represents a convenient way of determining
the onset of the pseudogap state.  (For definiteness, we define the
onset to correspond to $T^*=1.1T_c$). It is clear from the first two
figures that this occurs for 3d jellium at $g/g_c \approx 0.9$, and for
the quasi-2d case at $g/g_0 \approx 0.4$.\cite{g0=gc} This observation
reinforces the notion that pseudogap effects are easier to come by in
lower-dimensional systems. Similar behavior is seen in the quasi-2d
lattice situation for the $d$-wave case, although the energy scales on
the horizontal and vertical axes reflect the parameter $t_\parallel$
(rather than $g_c$ and $E_F$).\vspace*{-0.01in}

\section{Implications for the Cuprates}

There has been much concern in the literature about
whether generalized BCS Bose-Einstein crossover theories are relevant to
the copper oxide superconductors. Is the coupling $g$ sufficiently
``large" in some sense to warrant this form of departure from
conventional BCS theory?  In more concrete \linebreak

\noindent
terms, one may ask if the
calculated energy scales for $\Delta_{\text{pg}}$, $\mu$, $T^*$, and
$T_c$ are consistent with experiment? Are there other effects which are
more important than is the role of small $\xi$?  Perhaps among the most
intriguing questions raised is how does one incorporate hole
concentration (denoted by $x$) dependences into this picture?

An early motivation for adopting these crossover approaches was the
observed short coherence length $\xi$, which was suggestive of some
form of ``real space pairing.'' It is also clear that these systems are
doped Mott insulators\cite{Anderson87} so that the metal
insulator transition at 1/2 filling ($x=0$) should be integrated into
any theoretical approach. This transition is generally\cite{Lee2,Emery}
parametrized through an ``order parameter" such as the plasma
frequency $\omega_p$ which must necessarily van-
\linebreak

\begin{figure}
\vspace*{-0.32in}
\centerline{\leavevmode\hskip -0.4cm
\epsfxsize=3.in
\epsfbox{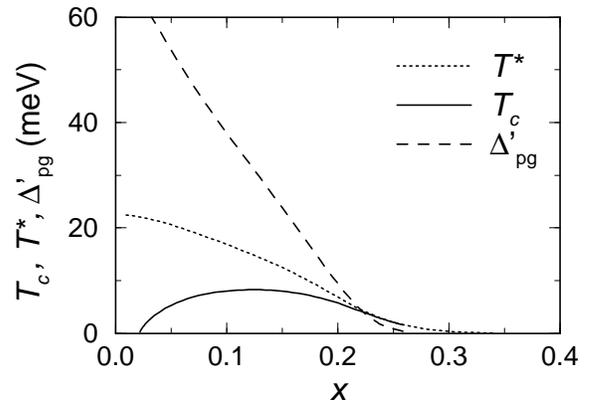}
}
\caption{Doping dependence of $T_c$, $T^*$, and
  $\Delta^\prime_{\text{pg}} [\equiv 2\Delta_{\text{pg}}$, the magnitude
  of the pseudogap at $(\pi, 0)$]. Here $t_\perp/t_\parallel=0.01,
  t_0=0.5$ eV, and $g$ is assumed constant, while the symmetry is
  $d$-wave.}
\label{Fig6}
\end{figure}

\noindent
ish as $x \rightarrow
0$.  Finally, it should be noted that there are no dramatic
effects on $\xi$ with variable $x$.\cite{Salamon}

It is clear that, in any attempt to understand pseudogap phenomena in
the cuprates, both the small size of $\xi$ and that of $\omega_p$
should be addressed on an equal footing. Early work by our
group\cite{Malycoulomb} investigated the effects of small $\omega_p$
on the crossover problem, at the level of the Nozi\`eres Schmitt-Rink
approximation, for charged fermions.  Coulomb interactions were
treated in the RPA in parallel with the RPA-like ladder diagrams of
the particle-particle attraction.  These early calculations
established that deviations from the BCS limit were more pronounced,
the smaller the plasma frequency. Thus, for the same value of $g$,
proximity to the insulating state is correlated with a tendency
towards ``bosonic" superconducting transitions. Additional effects (in
the same direction) result from the likely decrease in dimensionality
as the insulator is approached. Both of these effects, thus, suggest
an amplification of pseudogap phenomena as $x \rightarrow 0$.

In this section we address two issues which have been raised as
relevant for the cuprates: we examine the size of the various energy
scales for the case of $d$-wave superconductors (all of which depend
on the hopping matrix element $t_\parallel$), and we discuss some
aspects of the hole concentration dependence of the phase diagram,
beyond the very general qualitative issues which have been noted
above. It should be stressed that the fundamental basis of all
crossover theories is mechanism independent.  No information is used
or assumed about the details of the pairing mechanism beyond the
existence of a pairing coupling constant $g$. In almost any
microscopically based pairing scenario, $g$ is likely to contain some
degree of $x$ dependence. However, since there is no consensus on the
pairing mechanism, in the present paper it is inappropriate to obscure
our general results by making any detailed assumptions about the
nature of $g(x)$.

Here we focus exclusively on the $x$ dependence of the underlying
metal-insulator transition. We take $g$ as doping independent (which is
not unreasonable in the absence of any more detailed information) and
incorporate the Mott transition at half filling, by introducing an
$x$-dependence into the in-plane hopping matrix elements $t_\parallel$
of our calculations. In this way, we can explore the question of the
size of the various energy scales and capture some degree of hole
concentration dependence, albeit not the entire effect.  Our
renormalized band structure is based on the limit of extremely strong
on-site Coulomb repulsion, as seems appropriate for representing the
Mott transition. It follows from very early work on the Hubbard
model\cite{Anderson87} that the hopping matrix element is renormalized
as $t_\parallel(x) \approx t_0 (1-n)=t_0x$, where $t_0 (\approx 0.5
\text{eV})$ is the matrix element in the absence of Coulomb effects.
Equivalently, the effective particle mass varies as $1/x$. This change
of energy scale is consistent with the requirement that the plasma
frequency vanish at $x = 0 $.\cite{plasma}

In Fig.~\ref{Fig6} we replot the $d$-wave phase diagram of
Fig.~\ref{Fig4}b for the case of fixed $-g/4t_0=0.045$ (and
$t_\perp/t_\parallel=0.01$), which is chosen to fit the measured size
of the pseudogap at $T_c$ for extremely underdoped
cuprates.\cite{data-fit} Shown in the figure are $T^*$,
$\Delta_{\text{pg}}$, and $T_c$.  Agreement with experiment may or may
not be fortuitous since the coupling constant was assumed to be
independent of $x$.  Nevertheless, the energy scales appear to be
consistent with those measured
experimentally\cite{arpesanl,arpesstanford,arpesanl2,Miyakawa,Oda} and
the $x$ dependent trends are not inconsistent.\cite{variation}

 It should be stressed that the results shown in the figure are robust
consequences of our crossover theory. As a result of $d$-wave
symmetry, $T_c$ vanishes at moderately strong coupling. Moreover, this
maximal coupling is a fairly universal number (i.e., independent of
$n$) for a given $t_\parallel$, over the physical range of hole
concentrations ($x<0.3$). [See, e.g., Fig.~\ref{Fig4}(b)].  Once a linear
$x$ dependence is enforced in $t_\parallel$, near half filling, $T_c$
decreases naturally as $x$ decreases, and vanishes for extremely low
doping concentration.  This feature is insensitive to the detailed
parametrizations of the model.

\section{Conclusions}

In this paper we have applied a previously discussed\cite{Janko,Maly2}
BCS Bose-Einstein crossover theory to complex situations which are more
physically relevant than are our earlier studies of 3d $s$-wave jellium. In
this way, we have determined the effects of quasi-two dimensionality, of
periodic discrete lattices, and of a $d$-wave pairing interaction. This
crossover theory yields results which appear consistent with known physical
constraints and plausibility arguments. Thus, in particular, our strict 2d
calculations yield a sensible interpolation scheme (for $\mu$) with $T_c$
strictly zero. The effects of the lattice are consistent with earlier Monte
Carlo and other approaches, yielding a vanishing strong coupling limit for
$T_c$ associated with an increase (with $g$) of the effective mass $M^*$ of
the fermion pairs. Finally our $d$-wave studies reveal that, for this
symmetry, the superconducting bosonic regime is essentially never reached.
$T_c$ is suppressed to zero at moderate coupling constants, presumably
because of the lowered pair mobility due to the constraints imposed by
$d$-wave symmetry: the pair size cannot be reduced beyond the scale of a
lattice spacing. These features should be appended to other observations in
the literature\cite{Leggett} which note that in the strong coupling limit a
$d$-wave superconductor will not exhibit gap nodes. Indeed, it is sometimes
argued that this provides a ``proof" that the cuprates (which exhibit
explicit $d$-wave symmetry) cannot be in the bosonic regime. Our results
appear to make this case even more strongly, since we find that $T_c$ will
be zero whenever a $d$-wave system is in the preformed pair limit.

Our paper includes a brief discussion of the relevance to the copper
oxide superconductors, wherein we impose the simplest possible
ingredients of a Mott transition to arrive at some indications of hole
concentration dependence and characteristic energy scale parameters,
such as $T_c$, $T^*$, and the pseudogap amplitude
$\Delta_{\text{pg}}$. The numbers which emerge seem to be reasonably
consistent with experiment, although we have made no assumptions about
the origin or hole concentration dependence of the pairing
interaction. In this way, one may argue that these crossover scenarios
provide useful insights into the pseudogap state of the cuprates.

\acknowledgments

We would like to thank M. Norman and B. Gyorffy for useful
discussions.  This research was supported in part by the Science and
Technology Center for Superconductivity funded by the National Science
Foundation under Award No. DMR 91-20000.


\begin{references}
\bibitem{low-n}
In the limit of unphysically small $n < 0.09$ the superconducting bosonic
  regime can be accessed for certain couplings, since the lattice and jellium
  models then become indistinguishable, and the internal structure of the pairs
  (dominated by the pairing symmetry) becomes less important.
\bibitem{kz-cutoff} 
  We emphasize that a cutoff for $k_\perp$ is crucial for the quasi-two
  dimensionality of the system; otherwise, by rescaling $k_\perp
  \rightarrow (m_\perp/m_\parallel)^{1/2}\,k_\perp$, the system can be
  transformed into an isotropic 3d jellium model.
\bibitem{quasi-2d-note}
Strictly speaking, a very large cutoff $\Lambda$ for $k_\perp$ or a very low
  particle density (i.e., $k_F\ll \Lambda$) is necessary to restore the 3d
  limit if $m_\perp/m_\parallel=1$ is fixed. Alternatively, 3d can be achieved
  by letting $m_\perp/m_\parallel \rightarrow 0$ while fixing $\Lambda$. This
  complication does not occur for the lattice case, where all momenta are
  restricted within the first Brillouin zone.
\bibitem{zeroTc}
This is a high density effect at strong coupling. Lattice bandstructure effects
  and $d$-wave pairing serve to enhance this behavior. Morever, it should be
  noted that the pseudogap $\Delta_{\text{pg}}\propto [T_c/(m/M^*)]^{3/2}$
  remains finite as both $T_c$ and $m/M^*$ approach zero.
\bibitem{CDW}
In a negative U Hubbard model, there also exists a competition between
  superconductivity and charge-density wave ordering. This may not be relevant
  for a $d$-wave pairing interaction.
\bibitem{Itmaybereentrant}
In some cases we find $T_c$ may be reentrant, at larger $g$.
\bibitem{jellium+lattice} 
  One may argue that this quasi-2d system is not physical because the
  kinetic energy in the perpendicular direction should have a band
  structure, i.e., $2t_\perp(1-\cos k_\perp$) instead of being simply
  $k_\perp^2/2m_\perp$. In fact, it can be shown analytically that in
  this case, $M^*_\perp/M^*_\parallel\propto
  \left(m_\perp/m_\parallel\right)^2$ (instead of
  $m_\perp/m_\parallel\,$), i.e., the pairs see a magnified anisotropy.
  This does not affect the conclusion here.
\bibitem{pseudo-note}
The onset coupling $g$ for the pseudogap decreases when the dimensionality is
  reduced, and eventually vanishes in strictly 2d.
\bibitem{n-choice}
We show results for these values of $n$ because they corresponds to overdoped,
  optimally doped, and underdoped cuprates, respectively.
\bibitem{low-n-s} 
  As in the 3d ($s$-wave) lattice case, the bosonic regime can be
  accessed at $n < n_c \approx 0.53$ for sufficiently strong coupling.
\bibitem{Slatticephase}
  The phase diagram for the quasi-2d $s$-wave lattice case is similar to
  that of the quasi-2d $d$-wave case.
\bibitem{data-fit}
We take $\Delta_{\mbox{pg}} \approx 30$ meV [note $|\varphi_{\mathbf{k}}|=2$ at
  $(\pi, 0)$ in our convention] at $T_c$ for $x\approx 0.05$ and $t_0\approx
  0.5$ eV for definiteness.
\bibitem{variation}
There are some experimental variations from paper to paper. See Refs.
  \onlinecite{arpesanl,arpesstanford,arpesanl2,Miyakawa,Oda}.
\end{references}

\setlength{\textheight}{7.1in}

\setlength{\footskip}{2.5in}

\end{document}